\documentclass[prb,showpacs,aps,twocolumn]{revtex4-1}
\usepackage{graphicx}% Include figure files
\usepackage{epstopdf}
\usepackage{bm}% bold math
\usepackage{color}
\usepackage{ulem}
\usepackage{amsfonts,amsthm}
\usepackage{amsmath}
\usepackage{amssymb}

\begin{document}
\newcommand{\tr}[1]{\textcolor{red}{#1}}
\newcommand{\tm}[1]{\textcolor{magenta}{#1}}
\newcommand{\tb}[1]{\textcolor{blue}{#1}}
\newcommand{\tms}[1]{\textcolor{magenta}{\sout{#1}}}
\newcommand{\tbs}[1]{\textcolor{blue}{\sout{#1}}}
\newcommand{\tg}[1]{\textcolor{green}{#1}}
\newcommand{\tc}[1]{\textcolor[rgb]{0.0,0.5,0.5}{#1}}
\newcommand{\tcs}[1]{\textcolor[rgb]{0.0,0.5,0.5}{\sout{#1}}}
\newcommand{\avrg}[1]{\left\langle #1 \right\rangle}
\newcommand{\eqsa}[1]{\begin{eqnarray} #1 \end{eqnarray}}
\newcommand{\eqwd}[1]{\begin{widetext}\begin{eqnarray} #1 \end{eqnarray}\end{widetext}}
\newcommand{\hatd}[2]{\hat{ #1 }^{\dagger}_{ #2 }}
\newcommand{\hatn}[2]{\hat{ #1 }^{\ }_{ #2 }}
\newcommand{\wdtd}[2]{\widetilde{ #1 }^{\dagger}_{ #2 }}
\newcommand{\wdtn}[2]{\widetilde{ #1 }^{\ }_{ #2 }}
\newcommand{\cond}[1]{\overline{ #1 }_{0}}
\newcommand{\conp}[2]{\overline{ #1 }_{0#2}}
\newcommand{\nn}{\nonumber\\}
\newcommand{\cdt}{$\cdot$}
\newcommand{\bra}[1]{\langle#1|}
\newcommand{\ket}[1]{|#1\rangle}
\newcommand{\braket}[2]{\langle #1 | #2 \rangle}
\newcommand{\bvec}[1]{\mbox{\boldmath$#1$}}
\newcommand{\blue}[1]{{#1}}
\newcommand{\bl}[1]{{#1}}
\newcommand{\bn}[1]{\textcolor{blue}{#1}}
\newcommand{\rr}[1]{{#1}}
\newcommand{\bu}[1]{\textcolor{blue}{#1}}
\newcommand{\red}[1]{\textcolor{red}{#1}}
\newcommand{\fj}[1]{{#1}}
\newcommand{\green}[1]{{#1}}
\newcommand{\gr}[1]{\textcolor{green}{#1}}
\newcommand{\cyan}[1]{\textcolor{cyan}{#1}}
\newcommand{\grs}[1]{\textcolor{green}{\sout{#1}}}
\newcommand{\fry}[1]{\fcolorbox{red}{yellow}{#1}}
\newcommand{\frys}[1]{\fcolorbox{red}{yellow}{\sout{#1}}}
\newcommand{\fy}[1]{\fcolorbox{white}{yellow}{#1}}
\newcommand{\fys}[1]{\fcolorbox{white}{yellow}{\sout{#1}}}
\newcommand{\fb}[1]{\tm{\framebox[8.5cm]{\parbox{8.4cm}{#1}}}} 
\newcommand{\fbs}[1]{\tm{\framebox[8.5cm]{\parbox{8.4cm}{\sout{#1}}}}} 
\definecolor{green}{rgb}{0,0.5,0.1}
\definecolor{blue}{rgb}{0,0,0.8}
\definecolor{orange}{rgb}{1.0,0.5,0.0}
\preprint{APS/123-QED}

\title{
Variational Monte Carlo Method in the Presence of Spin-Orbit Interaction and
Its Application to Kitaev and Kitaev-Heisenberg Model
}
%\if0
\author{Moyuru Kurita$^1$}
\author{Youhei Yamaji$^2$}
\author{Satoshi Morita$^3$}
\author{Masatoshi Imada$^1$}
\affiliation{$^1$Department of Applied Physics, The University of Tokyo, Hongo, Bunkyo-ku, Tokyo, 113-8656, Japan \\
$^2$Quantum-Phase Electronics Center (QPEC), The University of Tokyo, Hongo, Bunkyo-ku, Tokyo, 113-8656, Japan\\
$^3$Institute for Solid State Physics, The University of Tokyo, Kashiwa, Chiba 277-8581, Japan}
%\affiliation{CREST, JST, Hongo, Bunkyo-ku, Tokyo, 113-8656, Japan.}
%\date{February 25, 2014}% It is always \today, today,
\date{\today}% It is always \today, today,

\begin{abstract}
We propose an accurate variational Monte Carlo method applicable in the presence of the strong spin-orbit interactions.
The algorithm is applicable even in a wider class of Hamiltonians that do not have the spin-rotational symmetry.
Our variational wave functions consist of generalized Pfaffian-Slater wave functions that involve
mixtures of singlet and triplet Cooper pairs, Jastrow-Gutzwiller-type projections, and quantum number projections.
The generalized wave functions allow
describing states including a wide class of symmetry broken states,
ranging from magnetic and/or charge ordered states to superconducting states and their fluctuations,
on equal footing without any {\it ad hoc} ansatz for variational wave functions.
We detail our optimization scheme for the generalized Pfaffian-Slater wave functions with complex-number variational parameters.
Generalized quantum number projections are also introduced, which imposes the conservation of not only the momentum quantum number but also Wilson loops.
As a demonstration of the capability of the present variational Monte Carlo method,
the accuracy and efficiency is tested for the Kitaev  and Kitaev-Heisenberg models,
which lack the SU(2) spin-rotational symmetry except at the Heisenberg limit.
The Kitaev model serves as a critical benchmark of the present method: The exact ground state of the model is a
typical gapless quantum spin liquid far beyond the reach of simple mean-field wave functions.
%far beyond the \tb{range applicable by} simple mean-field wave functions.
The newly introduced quantum number projections precisely reproduce the ground state degeneracy of the Kitaev spin liquids,
in addition to their ground state energy.
An application to a closely related itinerant model described by a multi-orbital Hubbard model with the spin-orbit interaction also shows promising benchmark results.
%}
The strong-coupling limit of the multi-orbital Hubbard model is indeed described by the Kitaev model.
Our framework offers accurate solutions
for the systems where strong electron correlation and spin-orbit interaction coexist.

\end{abstract}

%05.30.Rt Quantum phase transition
%71.10.Fd Hubbard model electronic structure
%73.43.Lp quantum Hall effects
%71.27.+a Strongly correlated electron systems

\maketitle

\section{Introduction}
In condensed matter physics, many fundamental and challenging issues are found in connection to strong correlation effects of electrons \cite{ImadaRMP}.
These are characterized by the interplay of itinerancy due to the kinetic energy and localization promoted by the Coulomb repulsions.
The interplay of the itinerancy and localization leads correlated electron systems to rich symmetry-broken phases such as magnetism \cite{Quantum_Magnetism} and high-$T_c$ superconductivity. \cite{Bednorz}
Even when electrons are localized due to dominant Coulomb repulsions, as realized in Mott insulators,
zero-point motions or quantum fluctuations of electronic spin and orbital degrees of freedom
induce competition among such symmetry-broken phases and quantum melting of them.
%\bn{[$\ast$ I could not understand what you want to say. The sentence, ``phase transitions bring rich physics such as
%magnetism and high $T_{\rm c}$,"
%has no information:
%Phase transitions, one of the main issue of modern physics, just occur in the point where the Coulomb interaction becomes dominant compared to the kinetic energy.
%They have brought many rich physics such as magnetism \cite{Quantum_Magnetism} and high-$T_c$ superconductivity \cite{Bednorz}.
%]}

Satisfactory understanding of the interplay and quantum fluctuations
requires sophisticated theoretical framework beyond the standard
band theory based on the one-body approximation.
However, correlated many-body systems are not rigorously
solvable except few cases\cite{Baxter,Lieb},
% in rigorous sense, 
while many numerical methods have been developed, and become powerful partly thanks to progress in the computational power.
Examples are exact diagonalization, auxiliary-field quantum Monte Carlo method\cite{Blankenbecler, Sorella, Imada, Furukawa}, density matrix renormalization group \cite{White}, and dynamical mean-field theory \cite{Metzner, Georges}.
Among all, the variational Monte Carlo (VMC) method\cite{Ceperley} offers a wide applicability after recent revisions as
we remark later.\cite{Bouchard, Sorella2001, Sorella2003, Bajdich, Tahara}
%\cyan{[References are not appropriate, we should cite papers on SR, Pfaffian wave functions, and Tahara,
 %Refs.\onlinecite{Bouchard, Sorella2001, Sorella2003, Bajdich, Tahara}]}.

Growing interest in topological states of matters urges the numerical methods
to be capable of handling strong relativistic spin-orbit interactions.
The topological states such as theoretically proposed topological insulators\cite{Kane_Mele, Moore, Roy, Hasan} followed by reports of several experimental realizations\cite{Konig, Zhang} have recently attracted much attention, where strong relativistic spin-orbit interaction plays crucial roles.
% are one of the most prominent examples, where the spin-orbit interaction plays a crucial role,
%and several experimental realizations \bn{have been} reported in this field \cite{Konig, Zhang}.
In addition, a certain class of orders with circular charge currents generates topologically non-trivial phases called Chern insulators\cite{Haldane,Regnault}.
Such circular-current orders realize non-collinear and non-coplanar alignments of local magnetic flux and affect electron dynamics.

Handling relativistic spin-orbit couplings and non-coplanar magnetism requires additional numerical costs and more complicated algorithm
in comparison with the conventional non-relativistic system with collinear magnetic orders.
Both of them break local symmetries and conservation laws such as SU(2)-symmetry of spins
and total $z$-components of spins.  
Furthermore, such complications generally make so-called negative sign problems more serious in 
quantum Monte Carlo methods.
For the understanding of correlated topological states of matters, %or further approach\bn{es} to the exotic phases,
it is important to construct a numerical framework to treat the spin-orbit interaction of electrons simultaneously with the strong Coulomb interaction.

In this paper, we formulate a generalized VMC method for the purpose of developing an efficient algorithm for systems without spin-rotational symmetry.
To test the efficiency of the method, we apply the method to the Kitaev and  Kitaev-Heisenberg models as well as to the multi-orbital Hubbard model with the spin-orbit interaction.
%\tr{\sout{.The part of the kinetic energy and the Coulomb interaction is already tractable for relatively large-sized systems.
%Here we further extend the VMC method to incorporate the spin-orbit interaction}}, 

The Kitaev model\cite{Kitaev} is a typical quantum many-body system that exhibits a 
topological state,
% example
%where 
which breaks both of SU(2)-symmetry of spins
and conservation of total $z$-components of spins.
There, a spin liquid phase becomes the exact ground state.
Though it is yet to be realized in experiments, %\cite{Liu}\cyan{[Is the reference appropriate ?]},
it is theoretically proposed that the Kitaev model is realized at strong coupling limit of the iridium compounds such as $\mathrm{{\it A}_2IrO_3}$ ({\it A}= Na, Li)\cite{Jackeli}, where the spin-orbit interaction plays a crucial role.

Due to their well-understood ground states and inherent strong spin-orbit couplings,
the Kitaev model and its variant, the Kitaev-Heisenberg model, are unique among various many-body systems.
They offer critical and suitable benchmark tests
of numerical algorithms for the topological states of correlated many-body systems under the strong spin-orbit interaction.

Though the VMC method inherently contains biases arising from the given form of wave functions,
the generalized Pfaffian-Slater wave functions together with quantum number projections
reduce the bias and considerably reproduce numerically exact results, which has partly been shown in the previous studies\cite{Tahara,Misawa}.
In this paper, we further show that our fermion wave functions give accurate description of the Kitaev liquid phase
with the help of newly introduced quantum number projections. The benchmark test on a multi-orbital Hubbard model with the spin-orbit interaction is also shown within the system sizes where the exact diagonalization result is available.
The multi-orbital Hubbard model is chosen as its strong-coupling limit is
described by the Kitaev model~\cite{Jackeli}.

The paper is organized as follows: We detail our VMC method including an energy minimization scheme and
quantum number projections in Sec. \ref{sec:VMC Method}.
In Sec. \ref{sec:App}, we apply our VMC method to the Kitaev and Kitaev-Heisenberg model.
In Sec. \ref{MultiOrbitalHubbard}, our method is applied to the multi-orbital Hubbard model with the spin-orbit interaction. 
Section \ref{sec:Discussion4} is devoted to discussion.

\section{VMC Method}\label{sec:VMC Method}

In this section, we present our VMC method.
We use pair wave functions with complex variables, where each pair of electrons is either singlet or triplet\cite{Bouchard,Sorella2003,Tahara}.
These variational wave functions are applicable to Hubbard and quantum spin models in the presence of the spin-orbit interaction and non-colinear magnetic fields
%\tr{When we consider these kinds of force, 
%\tr{\sout{In this extension, the} whose} 
whose Hamiltonian inevitably includes complex numbers,
such as
%$\bm{p} \times \bm{E}(\bm{r})$ 
spin-orbit couplings due to electric fields $\bm{E}(\bm{r})$, $[\bm{p} \times \bm{E}(\bm{r})]\cdot\bm{\sigma}$,
or the Pauli matrix $\sigma_y$.
%Therefore, the description of the resulting wave functions must include complex numbers which was not implemented in the real number VMC.
As a result, the formulation for the optimization becomes different from that with real numbers.

We consider the system which has $N_f$ degrees of freedom (including the number of site/momentum and flavors such as spins and orbitals)
and the wave function with $N$ fermions.
Then any many-body wave function is expanded in the complete set:
\begin{eqnarray}
|\psi_{\mathrm{full}} \rangle = \sum_{ \{ i_1 \cdots i_N\} } F_{ i_1 \cdots i_N }c_{i_1}^{\dagger}\cdots c_{i_N}^{\dagger}|0\rangle\cyan{,}
\end{eqnarray}
where $c_{i}^{\dagger}$ ($c_{i}^{\ }$) is a creation (an annihilation) operator with $i$-th degrees of freedom $(i=1,2,\cdots, N_f)$.
However, this function that expands the full Hilbert space and allows to describe the exact ground state is not tractable when the system size becomes large.
The pair-product function in the form
\begin{equation}
	|\psi\rangle = \left[ \sum_{i\neq j}^{N_f}F_{ij}c_{i}^{\dagger}c_{j}^{\dagger} \right]^{N/2}|0\rangle,
\end{equation}
instead was shown to provide us with an accurate starting point. 
In this work, we take $F_{ij}$ as complex variational parameters.
Here, the indices $i$ and $j$ specify the site, orbital, and spins.
Since anti-commutation relations of fermions require $F_{ji} = -F_{ij}$, there are $N_f(N_f-1)/2$ independent complex variational parameters and therefore $N_f(N_f-1)$ real variational parameters at maximum.
However, when the system has some symmetry and when we impose a related constraint to $F_{ij}$, this number can be reduced.
For example, when the translational symmetry is imposed on $F_{ij}$, the number of the variational parameters decreases to $N_u(N_f-1)$, where $N_u$ is the number of degrees of freedom contained in one unit cell.

Hereafter, independent real variational parameters are written as $\{\alpha_k | k = 1, \cdots, p\}$.
Here, index of pair-function variational parameter $k$ is ellipsis notation of $i,j$ as $F_{ij} = -F_{ji} = \alpha_{2l-1}+i\alpha_{2l}$.
%\gr{$\alpha_k$?ð??????‡???$F_{ij}$?ð??˜???õ??‡?ð??ð??}

We note that the pair function itself is a natural extension of the one-body approximation.
This is because when we have eigenstates of the one-body Hamiltonian as
\begin{eqnarray}
	H = \sum_{ij}t_{ij}c_{i}^{\dagger}c_{j} = \sum_{n=1}^{N_f}\epsilon_{n}c_{n}^{\dagger}c_{n}
\end{eqnarray}
and the $N$-particle eigenstate is described as
\begin{eqnarray}
|\psi\rangle = \prod_{n=1}^{N}c_{n}^{\dagger}|0\rangle,
\end{eqnarray}
the equivalent pair function is easily produced as
\begin{eqnarray}
	|\psi\rangle = \left[ \sum_{n=1}^{N/2}c_{2n-1}^{\dagger}c_{2n}^{\dagger} \right]^{N/2}|0\rangle,
\end{eqnarray}
where $c_{n}^{\dagger}$ is a creation operator of the $n$-th eigenstate of the one-body Hamiltonian.
In this calculation, we only used the fact that the Hamiltonian is one-body and it can contain any kind of spin-orbit interaction and magnetic field.
Therefore, the pair function of this form with complex variables can principally describe the state under spin-orbit interactions and non-colinear magnetic fields.

\subsection{Energy minimization}

To minimize the energy of the pair function with complex coefficient $F_{ij}$\cite{Weber}, we need to calculate
\begin{eqnarray}
	E = \frac{ \langle \psi_{\bm{\alpha}} | H | \psi_{\bm{\alpha}}\rangle}{ \langle \psi_{\bm{\alpha}}|\psi_{\bm{\alpha}}\rangle}, \ \mathrm{and} \ \frac{\partial E}{\partial \alpha_k},
\end{eqnarray}
with $H$ the Hamiltonian of the system.
Here, we explicitly wrote $|\psi_{\bm{\alpha}}\rangle$ to indicate that the wave function depends on variational parameters.
Since rigorous estimate of the energy is not possible if the system size becomes large, we apply the Monte-Carlo method.
For this purpose, it is convenient to introduce the normalized wave function $|\bar{\psi}_{\bm{\alpha}}\rangle$ and operators $O_k$ as
\begin{eqnarray}
|\bar{\psi}_{\bm{\alpha}}\rangle \equiv \frac{1}{\sqrt{\langle \psi_{\bm{\alpha}} | \psi_{\bm{\alpha}} \rangle }} | \psi_{\bm{\alpha}} \rangle,
\end{eqnarray}
and
\begin{eqnarray}
O_{k} &\equiv& \sum_{x}|x\rangle \left( \frac{1}{\langle x |\psi_{\bm{\alpha}} \rangle} \frac{\partial}{\partial\alpha_{k}} \langle x |\psi_{\bm{\alpha}} \rangle \right) \langle x | \notag \\
&=& \sum_{x}|x\rangle O_{k}(x) \langle x |.
\label{eq:Ok}
\end{eqnarray}
Here, $|x\rangle$ is real space configurations.
Then we get a expression for the derivatives of $|\bar{\psi}_{\bm{\alpha}}\rangle$ as
\begin{eqnarray}
|\bar{\psi}_{k\bm{\alpha}}\rangle &\equiv& \frac{\partial}{\partial \alpha_k} |\bar{\psi}_{\bm{\alpha}}\rangle \notag \\
&=& \frac{1}{\sqrt{\langle \psi_{\bm{\alpha}} | \psi_{\bm{\alpha}} \rangle }}\left[ O_k -\mathrm{Re}(\langle O_k \rangle ) \right] | \psi_{\bm{\alpha}} \rangle, \notag \\
&=& \left[ O_k -\mathrm{Re}(\langle O_k \rangle ) \right] | \bar{\psi}_{\bm{\alpha}} \rangle,
\end{eqnarray}
where a bracket $\langle A \rangle$ for an operator $A$ means $\langle \bar{\psi}_{\bm{\alpha}} |A| \bar{\psi}_{\bm{\alpha}} \rangle$.
Here, we have used the relations
\begin{eqnarray}
O_{k} |{\psi}_{\bm{\alpha}}\rangle &=& \frac{\partial |{\psi}_{\bm{\alpha}}\rangle}{\partial \alpha_k}, \notag \\
\frac{\partial}{\partial \alpha_k} \langle {\psi}_{\bm{\alpha}} |{\psi}_{\bm{\alpha}}\rangle &=& 2\mathrm{Re}\left( \langle {\psi}_{\bm{\alpha}}|O_{k}|{\psi}_{\bm{\alpha}}\rangle \right).
\end{eqnarray}
The energy gradient is now expressed as
\begin{eqnarray}\label{eq.g_k}
g_{k} &\equiv& \frac{\partial}{\partial \alpha_{k}} \langle \bar{\psi}_{\bm{\alpha}} |H| \bar{\psi}_{\bm{\alpha}} \rangle \notag \\
&=& 2\mathrm{Re}\left( \langle \bar{\psi}_{\bm{\alpha}} |H| \bar{\psi}_{k\bm{\alpha}} \rangle\right) \notag \\
&=& 2\mathrm{Re}(\langle H O_{k} \rangle) - 2\langle H \rangle \mathrm{Re}(\langle O_k \rangle).
\end{eqnarray}
To perform the steepest descent method, we only need the values of $g_{k}$.
However, this method is unstable compared to the stochastic reconfiguration (SR) method\cite{Sorella2001}.
The SR method requires the value of matrix $S_{kl}$ defined by
\begin{eqnarray}
\Delta_{\mathrm{norm}}^{2} &\equiv& \| | \bar{\psi}_{\bm{\alpha}+\bm{\gamma}} \rangle - | \bar{\psi}_{\bm{\alpha}} \rangle \| ^2\notag \\
&=& \sum_{k,l=1}^{p} \gamma_{k}\gamma_{l}S_{kl} + O(\gamma^3),
\end{eqnarray}
where $\bm{\gamma}$ is a small deviation from $\bm{\alpha}$.
Since the expansion of $| \bar{\psi}_{\bm{\alpha}+\bm{\gamma}} \rangle$ up to the first order is
\begin{eqnarray}
	| \bar{\psi}_{\bm{\alpha}+\bm{\gamma}} \rangle = | \bar{\psi}_{\bm{\alpha}} \rangle + \sum_{k=1}^{p}\gamma_{k} | \bar{\psi}_{k\bm{\alpha}} \rangle + O(\gamma^2),
\end{eqnarray}
we get
\begin{eqnarray}
\Delta_{\mathrm{norm}}^{2} = \sum_{k,l=1}^{p} \gamma_{k}\gamma_{l}\langle \bar{\psi}_{k\bm{\alpha}}  | \bar{\psi}_{l\bm{\alpha}} \rangle + O(\gamma^3)
\end{eqnarray}
and therefore
\begin{eqnarray}\label{eq.S_kl}
S_{kl} &=& \mathrm{Re}(\langle \bar{\psi}_{k\bm{\alpha}}  | \bar{\psi}_{l\bm{\alpha}} \rangle ) \notag \\
&=& \mathrm{Re}(\langle O_{k}^{\dagger}O_{l} \rangle) - \mathrm{Re}(\langle O_{k} \rangle)\mathrm{Re}(\langle O_{l} \rangle).
\end{eqnarray}
The SR method gives the updated variational parameter by
\begin{eqnarray}
\alpha_{k}^{\mathrm{new}} = \alpha_{k}^{\mathrm{old}} + \Delta \alpha_{k},
\end{eqnarray}
where the change from the initial value is given by
\begin{eqnarray}
\Delta \alpha_{k} = -\Delta t \sum_{l=1}^{p}S_{kl}^{-1}g_{l}.
\end{eqnarray}
Here, $\Delta t$ is a small constant, which is determined empirically.
Equations (\ref{eq.g_k}) and (\ref{eq.S_kl}) are different from the case with only the real parameters for $F_{ij}$\cite{Tahara}. 
However, techniques for the Monte-Carlo calculation of these quantities are the same.
We produce real space configurations
\begin{eqnarray}
	|x\rangle = c_{i_1}^{\dagger} \cdots c_{i_N}^{\dagger} |0\rangle
\end{eqnarray}
%\begin{eqnarray}
%	|x\rangle = c_{r_1}^{\dagger} \cdots c_{r_N}^{\dagger}|0\rangle
%\end{eqnarray}
and use the fact
\begin{eqnarray}
	\langle x | \psi \rangle &=& (N/2)! \mathrm{Pf} X,
   \label{eq:pfaffian}
\end{eqnarray}
where $X$ is $N\times N$ skew-symmetric matrix with the element
\begin{eqnarray}
	X_{kl} = F_{i_ki_l} - F_{i_li_k}.
\end{eqnarray}
%\begin{eqnarray}
%	X_{ij} &=& F_{r_ir_j} - F_{r_jr_i}.
%\end{eqnarray}
Then the expectation value of an operator $A$ is calculated by Monte-Carlo simulation as
\begin{eqnarray}
	A &\equiv& \frac{ \langle\psi | A | \psi\rangle}{ \langle \psi|\psi\rangle}
	=\frac{1}{\sum_{x'}\rho(x')}\sum_{x}\rho(x)\frac{\langle x |A|\psi \rangle}{\langle x| \psi \rangle},
\end{eqnarray}
where the weight for the Metropolis algorithm $\rho(x)$ is defined as
\begin{eqnarray}
	\rho(x) &\equiv& \left| \langle x | \psi \rangle \right|^2 . \notag
\end{eqnarray}
This method requires four times more computational operations than the VMC method with only real variational parameters, because of complex number calculations.

\subsection{Quantum-number projection}
In general, Hamiltonian often has several symmetries such as the translational symmetry, point group symmetry, and SU(2) spin-rotational symmetry.
However, a single pair function does not always satisfy them, which results in higher energy compared to the correct ground state.
%In such a case, variational wave function combined with the quantum-number projection constructed by transformation operators $T^{(n)}$ with weights $w_n$ is helpful.
In such a case, variational wave function combined with the quantum-number projection is helpful.
Let us consider the quantum-number projection constructed by transformation operators $T^{(n)}$ with weights $w_n$ as
\begin{equation}
 \mathcal{P}\equiv\sum_{n}\omega_n T^{(n)}.
\end{equation}
The form of our wave function is now
\begin{eqnarray}
 \mathcal{P}
	|\psi\rangle
	&=& \sum_{n}w_{n}T^{(n)}\left[ \sum_{i\neq j}^{N_f}F_{ij}c_{i}^{\dagger}c_{j}^{\dagger} \right]^{N/2}|0\rangle \notag \\
	&=& \sum_{n}w_{n}\left[ \sum_{i\neq j}^{N_f}\tilde{F}_{ij}^{(n)}c_{i}^{\dagger}c_{j}^{\dagger} \right]^{N/2}|0\rangle,
	\label{eq:QNprojection}
\end{eqnarray}
where $\tilde{F}_{ij}^{(n)}$ is calculated from $T^{(n)}$ and $F_{ij}$ in the following manner.
For the unitary operator $T$ defined by
\begin{eqnarray}
	Tc_{i}^{\dagger}T^{-1} = \sum_{k}\alpha_{ik}c_{k}^{\dagger},
\end{eqnarray}
the operated wave function is written as
%\begin{eqnarray}
%	T|\phi\rangle
%	&=& \left[ \sum_{ijkl}F_{ij}\alpha_{k}(i)\alpha_{l}(j)c_{T_{k}(i)}^{\dagger}c_{T_{l}(j)}^{\dagger} \right]^{\frac{N}{2}}|0\rangle \notag \\
%	&=& \left[ \sum_{i,j}\tilde{F}_{ij}c_{i}^{\dagger}c_{j}^{\dagger} \right]^{\frac{N}{2}}|0\rangle
%\end{eqnarray}
\begin{eqnarray}
	T|\psi\rangle
	&=& \left[ \sum_{ijkl}F_{ij}\alpha_{ik}\alpha_{jl}c_{k}^{\dagger}c_{l}^{\dagger}\right]^{N/2}|0\rangle
\notag \\
	&=& \left[ \sum_{k\neq l}\tilde{F}_{kl}c_{k}^{\dagger}c_{l}^{\dagger} \right]^{N/2}|0\rangle
\end{eqnarray}
with
%\begin{eqnarray}
%	\tilde{F}_{ij} = \underset{T_{k}(i') = i, T_{l}(j') = j}{\sum_{i'j'kl}}F_{i'j'}\alpha_{k}(i')\alpha_{l}(j').
%	\label{eq:Fijtrans}
%\end{eqnarray}
\begin{eqnarray}
	\tilde{F}_{kl} \equiv \sum_{i\neq j}F_{ij}\alpha_{ik}\alpha_{jl}.
	\label{eq:Fijtrans}
\end{eqnarray}
This type of projection is especially efficient when we consider a model
with SU(2) symmetry\cite{Tahara} or the Kitaev model
%\tr{\sout{The example with SU(2) symmetry is discussed in the previous study\cite{Tahara}}}, 
and the application to the Kitaev model will be discussed in the following section.

\section{Application to the Kitaev Model}
\label{sec:App}
\subsection{Model Hamiltonian}
\begin{figure}
\centering
\includegraphics[width=8.5cm]{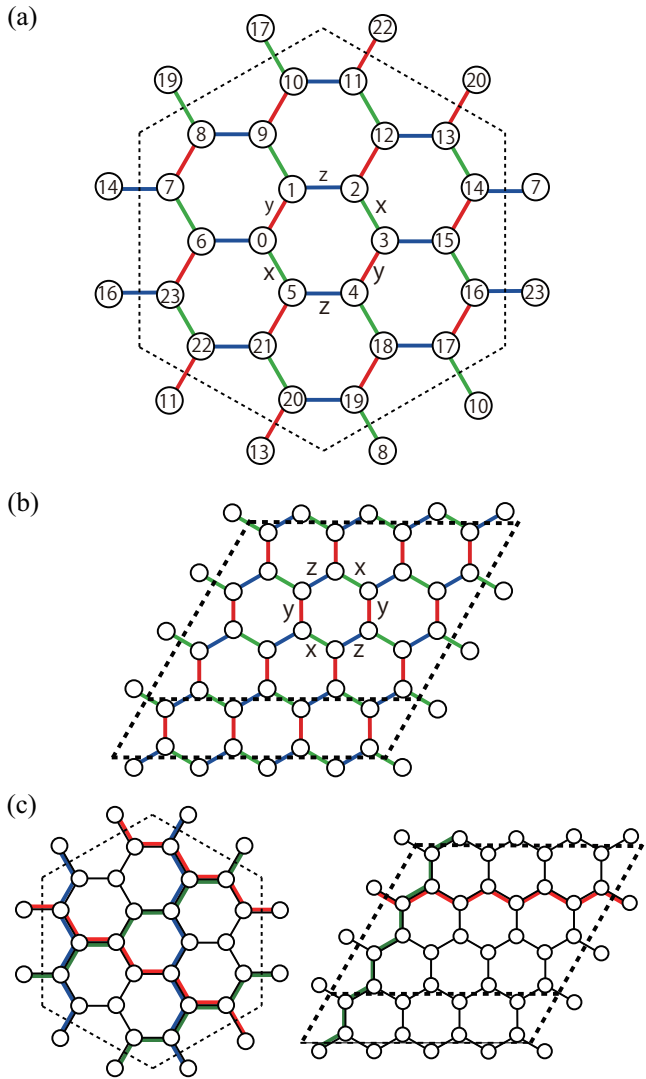}
\caption{
(Color online) (a) Honeycomb-lattice structure  on 24-site hexagonal supercell. Three types of bonds are illustrated as $x$ (green), $y$(red), and $z$ (blue) bonds, respectively.
%\tr{\sout{(Green, red, and blue) dashed lines are paths to define Wilson loop operators.}} 
(b) Honeycomb lattice with $3\times 4$ and $4\times 4$ unit cells contained in a supercell, where two configuration is indicated by two different (black) dashed boundary line.
In both pictures, dashed lines are periodic boundaries.
(c) Wilson-loop operators $W_x$, $W_y$, and $W_z$ defined along paths shown by green, red, and blue lines, respectively.
%\gr{$W_x$?ð??ð??ð??????ƒÎ?ð?????ð?????ðL?ð??ð??ð??ð??ð?????ð??ð??ð??ð??}}
}
\label{fig:fig1}
\end{figure}

In this section, we consider the Kitaev model and Kitaev-Heisenberg model.
By defining $S_{I\gamma} = \frac{1}{2}\bm{c}_{I}^{\dagger}\sigma_{\gamma}\bm{c}_{I}$, where $\bm{c}_{I}^{\dagger} = (c_{I\uparrow}^{\dagger}, c_{I\downarrow}^{\dagger})$ is a creation operator at the site $I$ and $\sigma_x, \sigma_y, \sigma_z$ are the Pauli matrices, the Kitaev \cite{Kitaev} and Kitaev-Heisenberg \cite{Chaloupka} Hamiltonian are described by
\begin{eqnarray}
	H_{\rm K} &=& -J_{{\rm K}x}\sum_{x-\mathrm{bond}}S_{Ix}S_{Jx} 
   - J_{{\rm K}y}\sum_{y-\mathrm{bond}}S_{Iy}S_{Jy} \notag \\
   &&- J_{{\rm K}z}\sum_{z-\mathrm{bond}}S_{Iz}S_{Jz},
\label{eq:Kitaev}
\end{eqnarray}
and
\begin{eqnarray}
	H_{\rm KH} &=& 2\alpha H_{\rm K}+ (1-\alpha) J_{\rm H}\sum_{\langle IJ \rangle} \bm{S}_{I} \cdot \bm{S}_{J},
	\label{eq:Kitaev-Heisenberg}
\end{eqnarray}
respectively.
Hereafter, we consider the case $J\equiv J_{{\rm K}x} = J_{{\rm K}y} = J_{{\rm K}z} = J_{\rm H}$ and take the energy unit $J=1$.
The characteristics of these Hamiltonians are that the anisotropy of the interaction depends on the bond direction.
There are three types of bonds ($x,y,z$-bond) as schematically shown in Fig. \ref{fig:fig1} corresponding to the bonds in Eq.(\ref{eq:Kitaev}).
%\end{document}

One interesting and important feature of the Kitaev Hamiltonian is the existence of local operators which commute with each other and with the Hamiltonian\cite{Kitaev}.
This conserved quantity is defined by the product of spin operators,
for example,
\begin{eqnarray}
K_{0} \equiv \Gamma_{0z}\Gamma_{1x}\Gamma_{2y}\Gamma_{3z}\Gamma_{4x}\Gamma_{5y},
\label{Koperatpr}
\end{eqnarray}
where the site indices $0-5$ are what we defined in Fig.\ref{fig:fig1}, and $\Gamma_{I\gamma} = 2S_{I\gamma}$.
These sites labeled by $0-5$ form a hexagon, and obviously there are local conserved quantities as many as the number of the hexagons in the system and they all mutually commute each other.
Actually, when we consider a honeycomb lattice with $N_s$ sites, there
are $N_q$ hexagons $(N_q= N_s/2)$ with corresponding operators $K_q$, and relations
\begin{eqnarray}
	[K_q,K_{q'}] = [K_q,H_{\rm K}] = 0
\end{eqnarray}
are satisfied. 

\subsection{Method and Result}
The VMC method precisely reproduces ground state wave functions
of quantum spin models with high accuracy by employing
the fermionic wave function of the form
\begin{eqnarray}
	|\psi_{\mathrm{pair}} \rangle = \mathcal{P}_\text{G}
   \left[ \sum_{i\neq j}^{N_f}F_{ij}c_{i}^{\dagger}c_{j}^{\dagger} \right]^{N/2}|0\rangle,
	\label{eq:VMCfunction1}
\end{eqnarray}
where $\mathcal{P}_\text{G}$ denotes the Gutzwiller projector defined as
\begin{equation}
 \mathcal{P}_\text{G} \equiv \prod_{I=1}^{N_s} (1-n_{I\uparrow}n_{I\downarrow}).
\end{equation}
Here, indices $i,j$ include both site $I,J$ and spin indices.
The number of species of fermion is $N_f = 2N_s$, where the factor 2 is from the spin degrees of freedom and the number of particles
in the wave function
is $N=N_s$.
The Gutzwiller projector excludes the double occupancy
and strictly keeps condition that there is one electron per site.
In more technical term, we only choose real space configuration $|x\rangle$ without double occupancy in the Monte-Carlo calculation.

We calculated three configurations of finite size honeycomb lattice as examples.
One is hexagonal 24 sites with periodic boundary condition as shown in Fig.\ref{fig:fig1} (a).
This configuration is exactly the same as that studied by Chaloupka\cite{Chaloupka}.
The others are supercells containing $3\times 4$ and $4\times 4$ unit cells under periodic boundary conditions with two-sublattice unit cell as shown in Fig.\ref{fig:fig1} (b).
One can compare results of these configurations with exact solutions and they are all close to the thermodynamic results.
We show comparisons of the VMC calculation with the exact diagonalization results.

We first show the results for the Kitaev model, where $\alpha=1$. The
variational parameters $F_{ij}$ are optimized by the SR method.
%\grs{In this first benchmark, we do not impose constraints on $F_{ji}=-F_{ij}$
%\cyan{in $|\psi_{\mathrm{pair}}\rangle$},
%but we
%do not operate any quantum number projection either.}
In this first benchmark, we do not impose any constraints on $F_{ij}$
except the anti-commutation relation, $F_{ji}=-F_{ij}$.
Thus the number of independent complex variational parameters in $|\psi_{\mathrm{pair}}\rangle$
is $N_f(N_f-1)/2$ as mentioned before.
The results are shown in the first row of Table \ref{Table1}.
As can be seen from the table, the errors in energy are as large as $5\%$
%\cyan{[$\ast\ast\ast$ too large !!]} 
in the $3\times4$- and $4\times4$-unit-cell boundary conditions.
This fact indicates that the wave function defined by Eq.(\ref{eq:VMCfunction1}) is insufficient to express the Kitaev liquid.

\begin{table}[t]
\caption{
Summary of the calculated total energy of Kitaev model. 
}
\begin{tabular}{c|c|c|cc}
\hline\hline
& hexagonal 24 sites & 3$\times$4 unit cells & 4$\times$4 unit cells \\
\hline
$|\psi_{\mathrm{pair}} \rangle$ & $-9.47 \pm 0.01$ & $-9.02 \pm 0.02$ & $-11.95 \pm 0.03$ \\
$|\psi_{\mathrm{MP}} \rangle$ & $-9.527 \pm 0.003$ &$-9.543 \pm 0.004$ & $-12.43 \pm 0.02$  \\
$|\psi_{\mathrm{KL}} \rangle$ & $-9.5285 \pm 0.0003$ & $-9.544 \pm 0.004$ & $-12.555 \pm 0.005$ \\
Exact & $-9.52860$ & $-9.54589$ & $-12.56409$ \\
\hline
\hline
\end{tabular}
\label{Table1}
\end{table}

We next study how the accuracy improves by imposing the quantum number
projection.
We consider two kinds of projections.
%The first is the total momentum projection by translational operators defined by
  % We first impose \gr{the} total momentum projection by translational operators defined by
The first is the total momentum projection (MP)
\begin{equation}
 \mathcal{P}_{\boldsymbol{K}=0} = \frac{1}{N_s} \sum_{\boldsymbol{R}} T_{\boldsymbol{R}},
\end{equation}
where the translational operator is defined as
\begin{eqnarray}
	T_{\bm{R}}c^{\dagger}_{I\sigma}T_{\bm{R}}^{-1} = c^{\dagger}_{I+\bm{R}, \sigma}.
\end{eqnarray}
%\grs{Again we do not impose constraints on $F_{ji}=-F_{ij}$.}
The resultant wave function has the form
\begin{eqnarray}
|\psi_{\mathrm{MP}} \rangle
 = \mathcal{P}_\text{G} \mathcal{P}_{\boldsymbol{K}=0}
 \left[ \sum_{i\neq j}^{N_f}F_{ij}c_{i}^{\dagger}c_{j}^{\dagger} \right]^{N/2}|0\rangle.
	\label{eq:VMCfunctionMP}
\end{eqnarray}
We note that the constraint on $F_{ij}$ in $|\psi_\text{MP}\rangle$ is the same as $|\psi_\text{pair}\rangle$.
%\gr{$F_{ij}$?ðL??Ý????ð?????ð??ð'?ð÷?ð?????ð??ð??}

The second projection is introduced by utilizing the fact that the eigenvalue of $K_{q}$ is either $1$ or $-1$ in the eigenstates of the Hamiltonian.
We define a projection operator $1\pm K_{q}$ to fix the eigenvalue of $K_{q}$ to $\pm1$, respectively.
Since all the eigenvalues of $K_{q}$ are unity in the ground state, we consider the state with $K_{q}=1$ for all hexagons.
The corresponding projection operator is defined as
\begin{equation}
 \mathcal{P}_{\text{KL}} \equiv \frac{1}{2^{N_q-1}}\prod_{q=1}^{N_q-1}\left(1+K_q\right).
\end{equation}
The number of transfomation operators in this projection is $N_q-1$, because there are $N_q$ hexagons in the system and an identity
$\prod_{q=1}^{N_q}K_{q}=1$ holds.
Since this projcetion operator commutes with the Gutzwiller projector,
the projected wave function is written as
\begin{eqnarray}
	|\psi_{\mathrm{KL}} \rangle = \mathcal{P}_\text{G}
   \mathcal{P}_\text{KL}
  \left[ \sum_{i\neq j}^{N_f}F_{ij}c_{i}^{\dagger}c_{j}^{\dagger} \right]^{N/2}|0\rangle.
	\label{eq:VMCfunction2}
\end{eqnarray}
%\noindent
%\fry{?ð??ð????????ð??????ƒ¶?ð??L????????ð??ð˜?ð??ð?????ð?????ð?????ð??ð????}
%\noindent
%\fry{ñ?÷????????‡?ð??˜??????'??‡?ð??????'??????????Ý?ñƒ¶?????‹?????ð?????ð?????ð??ð??ð????}
%\noindent
%\fry{(33)????ð??????ƒÎ???????????'?ð???'?‡"???'?ð??ð?????ð??ð??ð??ð??ð??ð˜?ð??ð?????ð????}
In fact, this wave function is able to represent the Kitaev liquid (KL)\cite{Baskaran},
in which long-range spin-spin correlations exactly vanish.
For example, we have
\begin{eqnarray}
\langle \psi_{\mathrm{KL}} | S_{0x}S_{ix} | \psi _{\mathrm{KL}} \rangle
&=&\langle \psi_{\mathrm{KL}} | K_{0}S_{0x}S_{ix}K_{0} | \psi _{\mathrm{KL}} \rangle \notag \\
&=&-\langle \psi_{\mathrm{KL}} | S_{0x}S_{ix} | \psi _{\mathrm{KL}} \rangle \notag \\
&=&0,
\end{eqnarray}
where $K_{0}$ is what defined in Eq.(\ref{Koperatpr}) and the site $i$ is far enough from the site $0$ that $S_{ix}$ commute with $K_{0}$ (See also Fig.\ref{fig:fig1}).
%\cyan{original?ð????????ð??ð??ð??}
We can also show $\langle \psi_{\mathrm{KL}} | S_{0z}S_{iz} | \psi _{\mathrm{KL}} \rangle = 0$
($\langle \psi_{\mathrm{KL}} | S_{0y}S_{iy} | \psi _{\mathrm{KL}} \rangle = 0$)
by using another hexagon operator $K_j$ which contains $\Gamma_{0x}$ or $\Gamma_{0y}$ ($\Gamma_{0z}$ or $\Gamma_{0x}$).
%\cyan{[$\ast\ast\ast$$\langle \psi_{\mathrm{KL}} | S_{0y}S_{iy} | \psi _{\mathrm{KL}} \rangle = 0$]}.}
%\fry{$S_{0z}$?ð?$S_{iz}$?ð??????'?ðL?ð˜?ð??ð÷???ñƒ¶??ð?????ð?????ð????}

The projection operator in $|\psi_{\mathrm{KL}} \rangle$ is treated by using the quantum-number projection method in Eq.(\ref{eq:QNprojection}) by expanding the operator as
\begin{eqnarray}
	\prod_{q=1}^{N_q-1}(1+K_{q}) = \sum_{\{n_q\}} \prod_{q=1}^{N_q-1} K_{q}^{n_q},
	\label{KitaevProjection}
\end{eqnarray}
where $n_q$ takes $+1$ or $0$.
The summation is over $2^{N_q-1}$ terms and it requires substantial computational cost.
In the calculation of supercells containing 3$\times$4 and 4$\times$4 unit cells each, we impose the translational symmetry of the Bravais lattice to $F_{ij}$ to reduce the computational cost.
The translational symmetry is not a serious constraint when the translational symmetry is preserved in the ground state as in spin liquid states.

%To reduce computational costs, we assume that $F_{ij}$ has translational symmetry.

As shown in Table \ref{Table1}, these projections largely improve the energy accuracy.
%\gr{$\psi_{\rm MP}$?ð?$\psi_{\rm KL}$?ð????????????õ?ð????}
While the error of $|\psi_{\rm MP}\rangle$ compared to the exact value is about $1\%$,
the error of $|\psi_{\rm KL}\rangle$ is $0.1\%$.
Therefore the projection of $K_q$ is very useful in improving the accuracy.

\begin{figure}
\centering
\includegraphics[width=8.5cm]{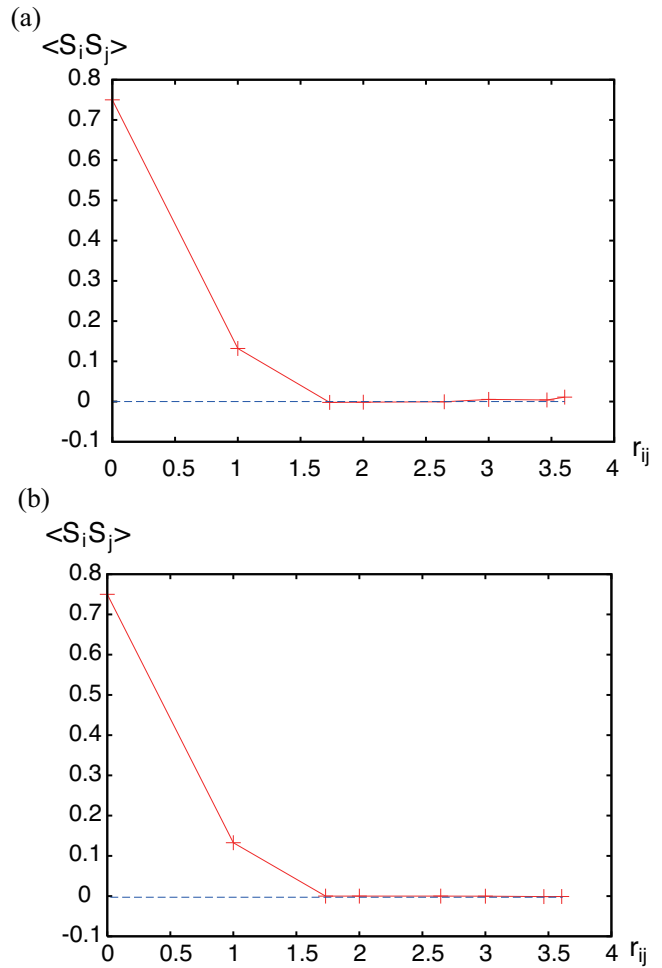}
\caption{
(Color online) Spin-spin correlations as functions of distances between $i$-th and $j$-th sites, $r_{ij}$ of (a) $|\psi_{\rm MP}\rangle$ and (b) $|\psi_{\rm KL}\rangle$ on $3\times 4$ unit cells.
Dashed (blue) lines show $\langle S_{i}S_{j}\rangle=0$.}
\label{fig:fig2}
\end{figure}

We also checked the accuracy of $|\psi_{\rm MP}\rangle$ and $|\psi_{\rm KL}\rangle$ by comparing the spin-spin correlations as a function of the distance with the exact result of the Kitaev model as shown in Fig. \ref{fig:fig2}.
These data are consistent with the exact result that only the nearest neighbor sites have nonzero spin-spin correlations.

\begin{figure}
\centering
\includegraphics[width=8.5cm]{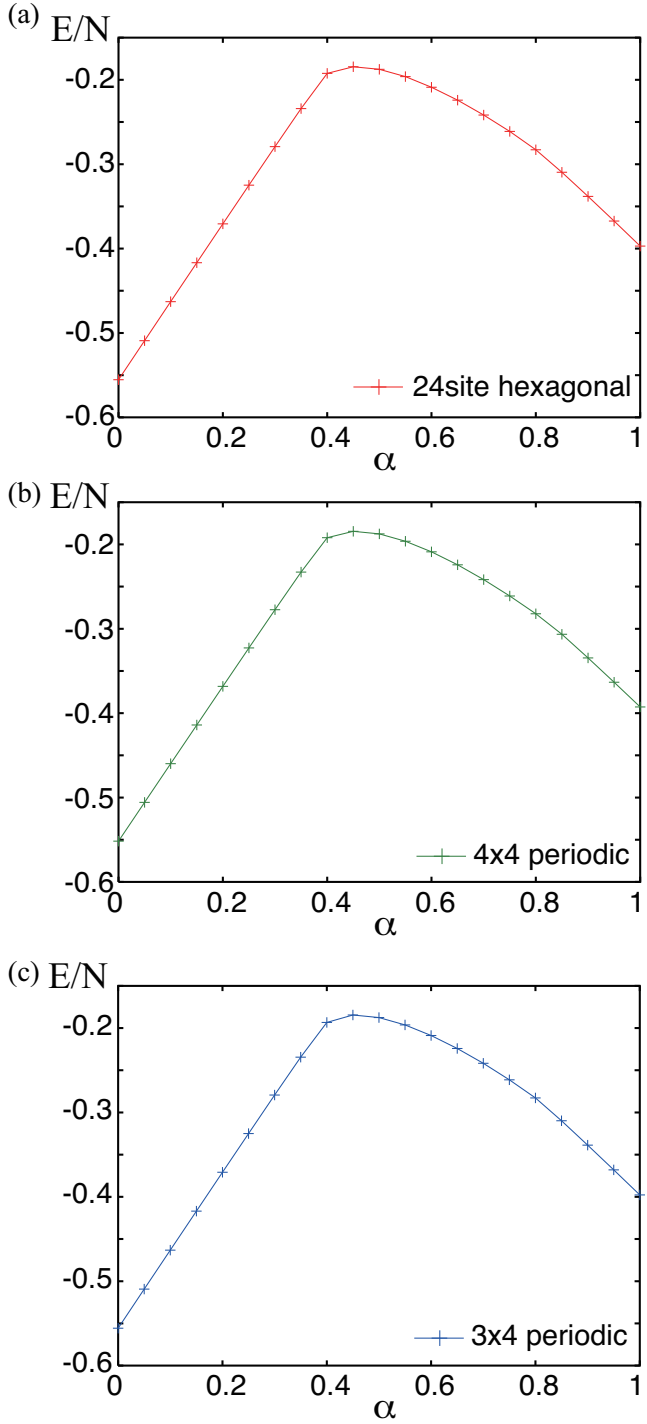}
\caption{(Color online)
Ground energies per site of the Kitaev-Heisenberg model calculated by exact diagonalization for (a) 24site hexagonal, (b) 3$\times$4 periodic, and (c) 4$\times$4 periodic, configuration.
The results show that the finite-size/lattice shape effects is very small.
}
\label{fig:fig3}
\end{figure}

\begin{figure}[h!]
\centering
\includegraphics[width=8.5cm]{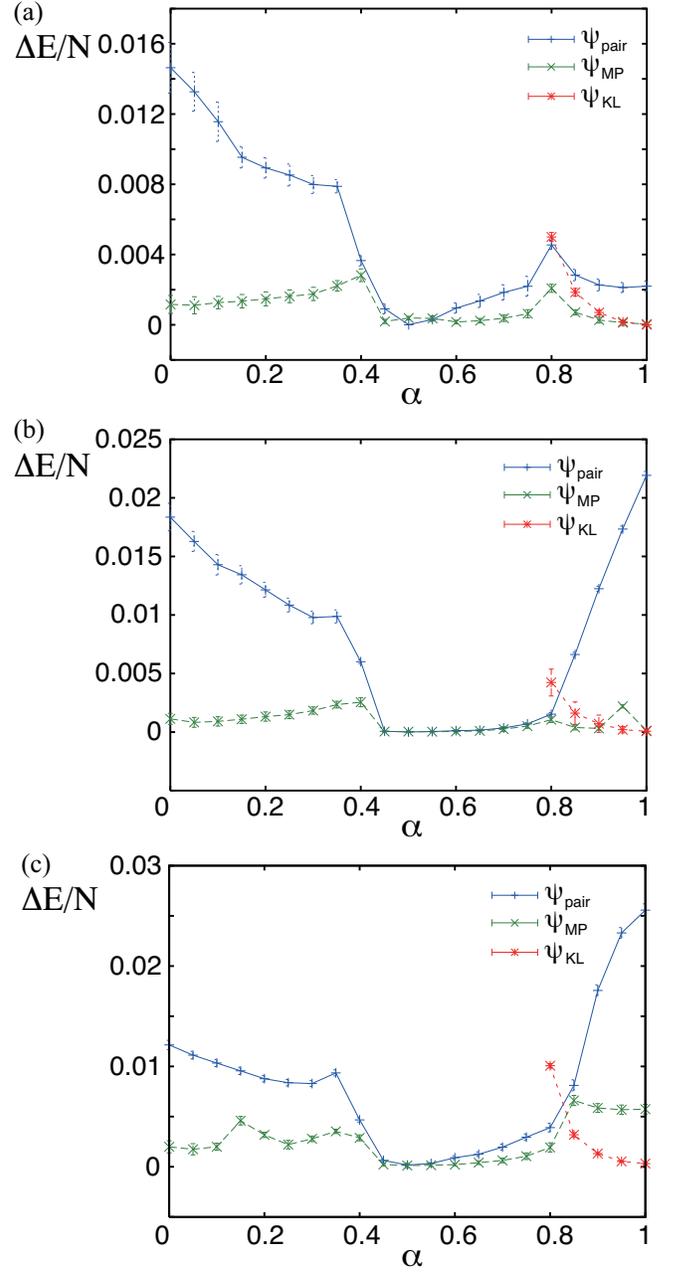}
\caption{(Color online)
Errors in energy per site for the VMC wave function in comparison to  the exact result calculated for hexagonal 24 sites lattice, $3\times4$-unit-cell lattice and $4\times 4$-unit-cell lattice respectively. Blue, green and red lines indicate the energy error of the VMC wave function without projection as in Eq.(\ref{eq:VMCfunction1}), with only momentum projection as in Eq.(\ref{eq:VMCfunctionMP}) and with only Kitaev liquid projection as in Eq.(\ref{eq:VMCfunction2}), respectively.
The error bars are estimated from the statistical errors of the Monte Carlo sampling and do not include the possible error in the SR optimization procedure.
}
\label{fig:fig4}
\end{figure}

\begin{figure}
\centering
\includegraphics[width=8.5cm]{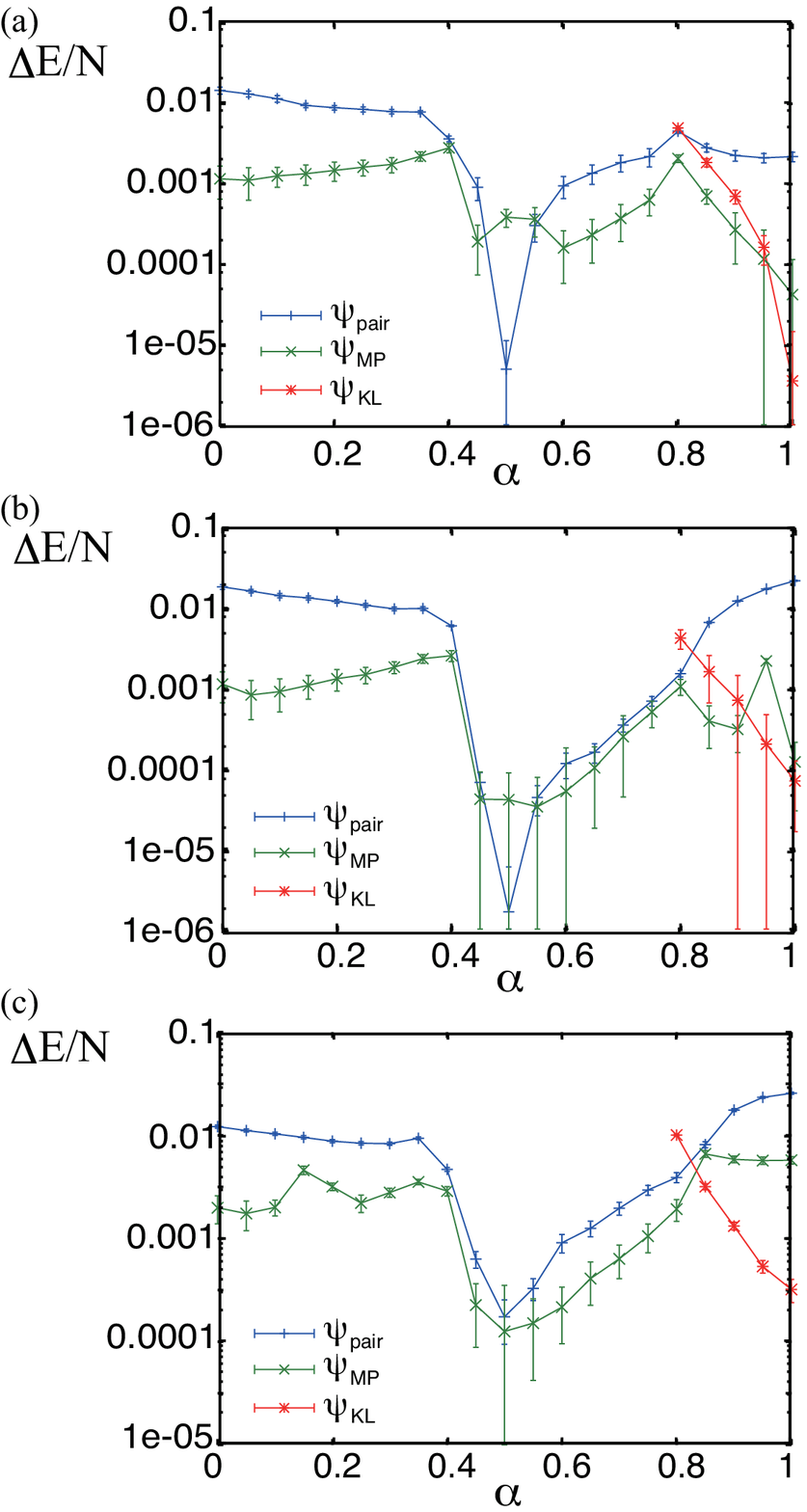}
\caption{(Color online)
Semi-log plot of errors in energy per site of the VMC wave functions shown in Fig.\ref{fig:fig4}.
Blue, green and red lines follow the same notations with Fig.\ref{fig:fig4}.
}
\label{fig:fig5}
\end{figure}

To further examine the applicability of our method,
we next study the Kitaev-Heisenberg model given by Eq.(\ref{eq:Kitaev-Heisenberg}).
The Heisenberg term is indeed naturally derived from the strong coupling expansion of the realistic fermionic Hamiltonian, which may coexist with the Kitaev term arising from the spin orbit interaction when the electron correlation and the spin orbit interaction are both strong
as in the case of 4$d$ and 5$d$ transition metal compounds.\cite{Chaloupka}
%\tr{The approach to the model which is not analytically solvable is quite important since when we consider much more realistic models, it necessarily contains some correction from the simple model.}
Figure \ref{fig:fig3} shows the ground energy of the Kitaev-Heisenberg model calculated by the exact diagonalization and Figs. \ref{fig:fig4} and \ref{fig:fig5} show the energy errors of the VMC results  using the function in Eqs.(\ref{eq:VMCfunction1}), (\ref{eq:VMCfunctionMP}) and (\ref{eq:VMCfunction2}) in comparison to the exact diagonalization.
A common property of these configurations is that the energy of the simple pair wave function $|\psi_\text{pair}\rangle$ becomes substantially higher than that with projection around and above $\alpha =0.85$.
Thus a simple pair function is inappropriate to describe the ground state of this region.
On the other hand, the error in the energy of the Kitaev liquid wave function $|\psi_\text{KL}\rangle$ becomes very small when $\alpha$ exceeds $0.9$, although hexagonal operators $K_r$ do not commute with the Hamiltonian $H_\text{KH}$ at $\alpha < 1.0$.
This indicates that as long as the ground state is adiabatically connected to the Kitaev limit, the function Eq.(\ref{eq:VMCfunction2}) works quite well as a variational wave function.
%The discrepancy of two methods becomes larger when $\alpha$ exceeds 0.8, and becomes largest at $\alpha = 1$.
%Indeed, it is known that the Kitaev-Heisenberg model shows phase transition to the Kitaev liquid phase at $\alpha \sim 0.8$ \cite{Chaloupka}.

The quantum number projection method is also useful to fix the eigenvalue of the topological Wilson loop operator which crosses the boundary.\cite{Kitaev}
Under the periodic boundary condition, two global loops are independent.
For example, in hexagonal 24-site lattice, two operators
\begin{eqnarray}
	W_{x} &=& \Gamma_{0x}\Gamma_{1x}\Gamma_{2x}\Gamma_{12x}\Gamma_{13x}\Gamma_{20x}\Gamma_{19x}\Gamma_{18x}\Gamma_{17x}\Gamma_{16x}\Gamma_{23x}\Gamma_{6x}, \notag \\
	W_{y} &=& \Gamma_{0y}\Gamma_{5y}\Gamma_{4y}\Gamma_{18y}\Gamma_{17y}\Gamma_{10y}\Gamma_{11y}\Gamma_{12y}\Gamma_{13y}\Gamma_{14y}\Gamma_{7y}\Gamma_{6y},\notag \\
\end{eqnarray}
have eigenvalues $\pm1$ and commute with the Kitaev Hamiltonian and $K_{q}$, where the paths of global loops are shown in Fig.\ref{fig:fig1}(c).
Therefore, the eigenstates of the Kitaev model are characterized by these quantum numbers.
Other Wilson loop operators can be decomposed into product of $W_x$, $W_y$ and $K_q$.
We note that corresponding $z$-spin Wilson loop operator in the other direction of the global loop 
\begin{eqnarray}
	W_{z} &=& \Gamma_{2z}\Gamma_{3z}\Gamma_{4z}\Gamma_{18z}\Gamma_{19z}\Gamma_{8z}\Gamma_{7z}\Gamma_{6z}\Gamma_{23z}\Gamma_{22z}\Gamma_{11z}\Gamma_{12z}, \notag \\
\end{eqnarray}
is not independent of $W_{x}$ and $W_{y}$, because $W_{x}W_{y}W_{z}=-1$ is satisfied in the Kitaev liquid wave function owing to the identity
\begin{eqnarray}
W_{x}W_{y}W_{z}\prod_{q=1}^{N_q}K_{q} = -1.
\end{eqnarray}
%\gr{????????'?ð?????ð??ð'?ð÷?ð?????ð??ð????$(W_x, W_y)$?ð??ð??ð?????‡?????????L?????????????õ??'??ðL????ð??ð÷???????ð??ð÷?ð???????VMC?ðL??????????????????????˜ç?ð??ð÷?ð??ð??ð??ð?}
These facts are equally true in $3\times 4$ and $4\times 4$ unit-cells supercell.
%\grs{Though they do not have the corresponding $W_z$ loop,} \gr{???????'??????‹ñ?Ý?ð?????ð‡$\sigma_z$?ð?????ð??‹?????ð??ð????Wilson loop?ð?????ð????????ð??ð???????}
$W_x$ and $W_y$ of these lattices are also shown in Fig.\ref{fig:fig1}(c).

The quantum numbers $W_x$ and $W_y$ have primary importance in understanding the topological order of the ground state of the model, because the four combinations of the Wilson loop numbers $W_x=\pm1$ and $W_y=\pm1$ characterize the four-fold degeneracy of the ground state in the thermodynamic limit.  Namely, the Wilson-loop number specifies the topological order.\cite{Kitaev}

Table~\ref{Table2} shows several exact results on low energy states for our choices of the lattice with eigenvalues of the Wilson loop operators for the finite-size systems we studied.
%\fb{4??????????ð÷???????ð??ðƒÎ?ð??ð??ð?????ð??ð?VMC?ð?????????'?ð????????ð??ð÷?ð?????ð??ð????}
\begin{table}[t]
\caption{
Comparisons of total energies of low-energy states obtained by the VMC with the exact diagonalization
for the three choices of finite lattice with corresponding values of the Wilson loop operators.
$+$ and $-$ are $1$ and $-1$ respectively.
}
\begin{tabular}{c|c|c|c}
\hline\hline
Configuration & ($W_x, W_y$) & Exact & VMC\\
\hline
hexagonal 24 sites & $(+, +)$ & $-9.5286$ &-9.528$\pm$0.002\\
& $(+, -)$ & $-9.5286$ &-9.528$\pm$0.002\\
& $(-, +)$ & $-9.5286$ &-9.528$\pm$0.002\\
\hline
3 $\times$ 4 unit cells & $(+, -)$ & $-9.5459$ & -9.546$\pm$0.001\\
%& $(+, +)$ & $-9.4934$ \\
\hline
4 $\times$ 4 unit cells & $(+, +)$ & $-12.5641$ &-12.55$\pm$0.01\\
&$(+, -)$ & $-12.5641$ &-12.55$\pm$0.01\\
&$(-, +)$ & $-12.5641$ &-12.55$\pm$0.01\\
\hline
\end{tabular}
\label{Table2}
\end{table}

In fact, our VMC result by $|\psi_\text{KL}\rangle$ supplemented by the Wilson loop projection shows that the ground state of 24 site hexagonal configuration is triply degenerate classified by these Wilson loop operators.
When we define three wave functions
\begin{eqnarray}
	| \psi_{x} \rangle &=& (1-W_{x})(1+W_{y})|\psi_{\mathrm{KL}} \rangle \notag \\
	| \psi_{y} \rangle &=& (1+W_{x})(1-W_{y})|\psi_{\mathrm{KL}} \rangle \notag \\
	| \psi_{z} \rangle &=& (1+W_{x})(1+W_{y})|\psi_{\mathrm{KL}} \rangle,
	\label{xyzbasis}
\end{eqnarray}
their energies go to the ground state energy as $-9.528\pm 0.002$ in excellent agreement with the 
exact results. %\fb{$W_x=W_y=-1$?ð?VMC?ð???????????ð?????????L???????ð??ð??ð??ð????}

Since the point group symmetry of the Kitaev model and
the size of its largest irreducible representation is two (see also Appendix A),
the triple degeneracy of the ground states is not of the point group symmetry and inevitably must have a topological origin characterized by the Wilson-loop quantum number.
As we mentioned above, in the thermodynamic limit (on the infinite size), the topological order of the Kitaev model predicts the four fold degeneracy. However, the convergence of the state with $W_x=W_y=-1$ to the ground state resulting in the four-fold degeneracy seems to be slow with the increase in the system size.

We also note that when we fix eigenvalues of Wilson loop operators, in
the $4 \times 4$ unit-cells lattice, the energies calculated with the
three conditions $(W_x, W_y) = (1,-1), (-1,1),(-1,-1)$ become as low as
$-12.55 \pm 0.01$.  In the case of the $3 \times 4$ unit-cells lattice,
it is different and we found the energy of $(W_x, W_y) = (1,-1)$ state
is $-9.546\pm0.001$.
% and $(W_x, W_y) = (1,1)$ state is $-9.546 \pm 0.001$.
Indeed the exact ground state of the $4 \times 4$ unit-cells lattice is
triply degenerate while that of the $3 \times 4$ is not degenerate.
The ground state energy of the latter is $-9.5459$ with $(W_x, W_y)=(1, -1)$ and its first excited energy is $-9.4934$ with $(W_x, W_y)=(1, 1)$.
All are consistent with our results.  Since the energy per
site is well converged as $-0.3970$ (hexagonal 24 sites), $-0.3977$
(3$\times$4 lattice) and $-0.3926$ (4$\times$4 lattice) in comparison to
the thermodynamic value\cite{Kitaev}, $-0.39365$, we expect that the present method
keeps the accuracy and efficiency for increased system size.

%\fb{ñ??ñ???ð??????‡?ð???ðñƒ¶????????ð??ð??ð??ð??ð?????ð?????ð??ð?. ñ??ñ???ð??????ƒ¶?ð?????÷ð?ð????}
%\begin{eqnarray}
%&&T\left[ \sum_{i\neq j}^{N_f}F_{ij}c_{i}^{\dagger}c_{j}^{\dagger} \right]^{\frac{N}{2}}|0\rangle \notag \\
%&=& \left[ \sum_{i\neq j}^{N_f}F_{ij}Tc_{i}^{\dagger}T^{-1}Tc_{j}^{\dagger}T^{-1} \right]^{\frac{N}{2}}|0\rangle.
%\end{eqnarray}
When the system size becomes larger, it is expected that one excited level becomes lower and we get four fold degeneracy in the infinite size limit (see also Appendix B). 
%\tm{?ð??ð??ð??ð÷?ð??ð?ñ??????ð??ð?????ð??õƒÎ??õ?ð?????ð??ð÷?ð??ð????}
%\fb{?ð??ð??????‹???????ð??ð?????'??ð??÷??ƒ¶??ð?????ð??ð????}
%\cyan{?ð??ð'?ð??ð???‡?ð??ð?????ð??ð÷?ð?????ð??ðõ?ð??????ðEq.(21)?ð?ñ???ð????????ð??ð?????ð??ðL??ð$T(I)$?ð?
%?????????????ð??ð???ƒÎ?ð?????ð????$\hat{T}$?ð??õ??'???ƒÎ?ðL????ð???????????ðõ?ð?????ð????????ð??ð?????ð????}

\begin{figure}
\centering
\includegraphics[width=8.5cm]{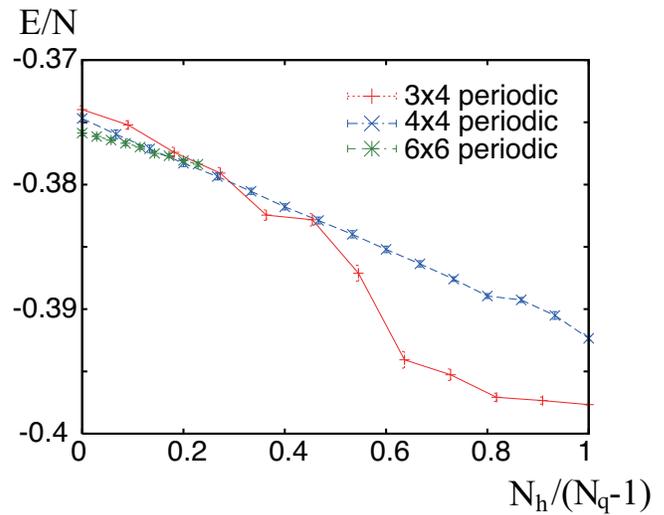}
\caption{(Color online)
Energy per site of VMC wave function at $3\times 4$, $4\times 4$, and $6\times 6$ unit cells configuration for different values of $N_h$ for the Kitaev model.}
\label{fig:fig6}
\end{figure}

We also note that our projection method efficiently improves the wavefunction by lowering energy even if it is a partial projection.
This enables us to calculate the wave function for large systems.
By replacing Eq.(\ref{KitaevProjection}) with
\begin{eqnarray}
	\sum_{\{n_q\}} \prod_{q=1}^{N_h} K_{q}^{n_q},
\end{eqnarray}
with a number of hexagonal operators $K_q$,
$N_{h}$, instead of $N_{q}-1$ where $N_h<N_{q}-1$, the computation time can be reduced to a tractable range.
In this projection, we arbitrarily choose some hexagons in the lattice to fix the value of $K_{i}$ to $1$.
The energy gets lower  for larger $N_h$ with monotonic reductions as shown in Fig. \ref{fig:fig6}, where the energies of $3\times 4$, $4\times 4$, and $6\times 6$ unit cells are calculated by the VMC.
The energies at $3\times 4$, $4\times 4$ converge accurately in the limit $N_h \rightarrow N_q-1$ as already shown in Table \ref{Table2}.
For the size $6\times 6$, although the full projection is not possible within our feasible computation time, the improvement of the energy with increasing $N_h$ suggests that it approaches the exact value $E/N = -3.901$.
Beyond the full summation, Monte Carlo sampling of the summation of projection discussed in Sec. \ref{sec:Discussion4}
may relax the limitation on $N_h$ and give convergence of energy within feasible computational time.

\section{Application to multi-orbital Hubbard model}
\label{MultiOrbitalHubbard}

We also note that our method is applicable to
Hamiltonians with explicit spin-orbit interactions.
Here we apply the present method to
a multi-orbital Hubbard model with spin-orbit interactions, which is closely related to the Kitaev model.
It is known that the Kitaev model is
effectively realized as strong-coupling limits of a
$t_{2g}$ multi-orbital Hubbard model on a honeycomb lattice\cite{Jackeli,Chaloupka,Kimchi,Katukuri,Rau2014,Yamaji}.
There, the hopping amplitude $t$, spin-orbit interaction $\lambda$, on-site Coulomb interaction $U$, and Hund's coupling $J_{H}$ are basic parameters to define the Hamiltonian.
The Hamiltonian is given by
\begin{eqnarray}
	H_{t2g}=H_{0}+H_{\rm SOC} + H_{U},
\end{eqnarray}
where each term is defined by the following:
The hopping term is given by
\begin{eqnarray}
H_{0} = 
\sum_{\langle i,j \rangle \sigma}
\sum_{a,b=xy,yz,zx}
t_{i,j;a,b}
\left[
c^{\dagger}_{i a \sigma}c_{j b \sigma} + 
{\rm h.c.} \right].
\end{eqnarray}
Here we define the hopping matrices $t_{i,j;a,b}$
on the honeycomb lattice (see Fig.7) as follows:
If $\langle i,j\rangle$ belongs to a $x$-bond, $(a,b) = (zx,xy), \langle i,j \rangle \in y$-$\mathrm{bond}\ (a,b) = (xy,yz),$ and $\langle i,j \rangle \in z$-$\mathrm{bond}\ (a,b) = (yz,zx)$ and $0$ for others.
The spin-orbit interaction term is defined by
\begin{eqnarray}
H_{\rm SOC} = \lambda
\sum_{i}
\vec{c}^{\dagger}_{i}
\left[
\begin{array}{ccc}
0 & +i\hat{\sigma}_z & -i\hat{\sigma}_y \\
-i\hat{\sigma}_z & 0 & +i\hat{\sigma}_x \\
+i\hat{\sigma}_y & -i\hat{\sigma}_x & 0 \\
\end{array}
\right]
\vec{c}_{i},
\end{eqnarray}
where a vector representation $\vec{c}^{\dagger}_{i}=
(
c^{\dagger}_{\ell yz\uparrow},c^{\dagger}_{i yz\downarrow},
c^{\dagger}_{\ell zx\uparrow},c^{\dagger}_{i zx\downarrow},
c^{\dagger}_{\ell xy\uparrow},c^{\dagger}_{i xy\downarrow}
)$ is introduced.
Finally the two-body interaction term is defined by
\begin{eqnarray}
&&H_{U}
=U\sum_{i}\sum_{a=yz,zx,xy}n_{i a\uparrow}n_{i a\downarrow}
\nn
&&+
\sum_{i }\sum_{a<b}\sum_{\sigma}
\left[
(U-2J_{\rm H}) n_{i a\sigma}
{n}_{i b\overline{\sigma}}
% \hat{n}_{i b\overline{\sigma}}
+
(U-3J_{\rm H}) n_{i a\sigma}n_{i b\sigma}
\right]
\nn
&&
+J_{\rm H}\sum_{i }\sum_{a\neq b}
\left[
c^{\dagger}_{i a\uparrow}
c^{\dagger}_{i b\downarrow}
c_{i a\downarrow}
c_{i b\uparrow}
+
c^{\dagger}_{i a\uparrow}
c^{\dagger}_{i a\downarrow}
c_{i b\downarrow}
c_{i b\uparrow}
\right].
\end{eqnarray}

We defined the VMC wave function of hole picture to reduce the size of the matrix $X$ in Eq.(\ref{eq:pfaffian}) and introduced a Jastrow-Gutzwiller-type correlation factor of the form
\begin{eqnarray}
C_\text{JG} \equiv \exp\left( - \sum_{i,j} g_{ij}N_{i}N_{j} \right),
\end{eqnarray}
where $N_{i}$ is the number of hole at site $i$ and $g_{ij}$ are variational parameters.
Our VMC wave function is given by $C_\text{JG}|\psi_{\mathrm{pair}}\rangle$.
Calculating $O_{k}(x)$ in Eq.(\ref{eq:Ok}) for these parameters are not difficult.
For any type of correlation factor of the form
\begin{eqnarray}
C_{\alpha} = \exp\left( -\sum_{k}\alpha_{k}\Theta_{k} \right),
\end{eqnarray}
where $\alpha_k$ are variational parameters and $\Theta_{k}$ are diagonal operators for real space configurations $\left[ \Theta_{k}|x\rangle = \Theta_{k}(x)|x\rangle \right]$, corresponding $O_{k}(x)$ is obtained as
\begin{eqnarray}
O_{k}(x) = - \Theta_{k}(x).
\end{eqnarray}

For the benchmark calculation, we set the value to $t=\lambda = 1.0$, $U=2.5$ and $J_{\mathrm{H}} = 0,\ 0.25$ with the $5/6$-filled electron density ($5N_s$ electrons and $N_s$ holes for the $N_s$-site system).
In the case of a single hexagon with the periodic boundary condition, shown in Fig. \ref{fig:KHhexagon}, we found that the error of the energy calculated by our VMC method compared to that obtained by the exact diagonalization is about $0.1\%$ for this parameter values.
Though we can not obtain the exact energies for larger sizes in this Hamiltonian, 
the result suggests that the present method is also efficiently applicable to the itinerant models such as the multi-orbital Hubbard model with the spin-orbit interaction. Applications to the itinerant systems with the interplay of the electron correlation and the spin-orbit interaction beyond the benchmark is an intriguing future issue.

\begin{figure}
\centering
\includegraphics[width=8.5cm]{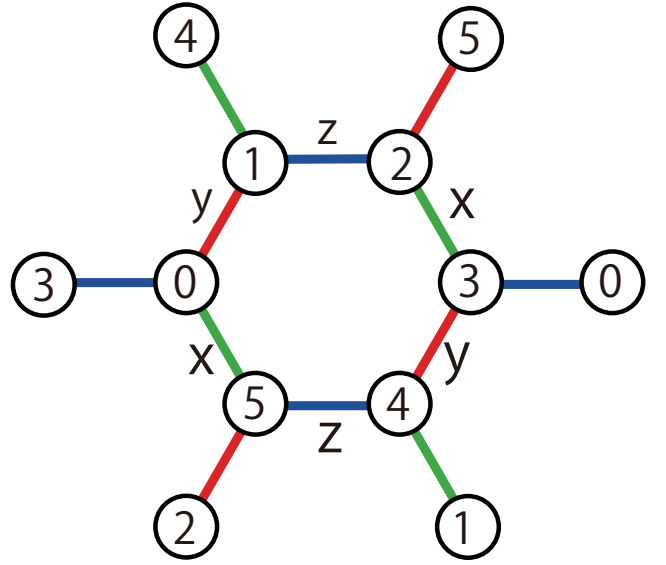}
\caption{(Color online)
Single hexagon with periodic boundary condition.
Here the size is $N_s=6$ with six degrees of freedom at each site (three orbitals times two spin degrees of freedom).
}
\label{fig:fig7}
\end{figure}

\begin{table}[t]
\caption{
Total energy of three-orbital Hubbard model with spin-orbit interaction $\lambda$ at $t=\lambda=1.0$, and $U=2.5$ . 
}
\begin{tabular}{c|c|c}
\hline\hline
$J_{\mathrm{H}}$ & VMC & Exact \\
\hline
0.0 & $ 135.85 \pm 0.02$ & 135.78 \\
0.25 & $ 105.64 \pm 0.03$ & 105.52 \\
\hline
\hline
\end{tabular}
\label{Table3}
\end{table}

\section{Discussion and Conclusive Remarks}\label{sec:Discussion4}
In this paper, we have extended the VMC method to treat the Hamiltonian, and applied it to the Kitaev-Heisenberg model.
The advantage of the VMC method is that it is able to treat large system sizes even when electron correlations and geometrical frustrations are large.
Our study further shows that the method is efficiently applicable to systems including non-colinear magnetic fields and spin-orbit interactions.
Furthermore, we have shown that the energy error in the Kitaev limit is very small (less than $0.1\%$ in the present case) without an appreciable system size dependence.
The degeneracy of topological ground state in the Kitaev limit is also successfully reproduced by employing
the quantum number projections, which is a powerful tool for studying Kitaev spin liquid. 

Reducing computational costs for calculations of large-size systems is one of the future issue.
In the present method, the fully projected wave function in the Kitaev spin model is tractable in limited sizes in practice because of the exponentially increasing number of summation for the projection by $K_q$ and sizes far beyond the present results are not practically feasible. However, a sampling of the projection will solve this difficulty in the future as we details below.

One of the future issue is to use the Monte-Carlo method in performing quantum number projections.
A certain class of quantum number projections including $\prod_{q=1}^{N_q-1}(1+K_q)$ requires
demanding computational costs exponentially scaled by the system size,
as already discussed below Eq.(31). Therefore, for larger system sizes, efficient algorithm to calculate
such quantum number projections is useful.
That is, for %\gr{ñ??????ðL?ð?$w_n$??????????ð÷?ð??ð??ðƒÎ?ð??ð??ð??}
\begin{eqnarray}
	|\psi\rangle &=& \sum_{n}w_nT^{(n)} |\phi\rangle,
\end{eqnarray}
physical quantity is expressed as
\begin{eqnarray}
	\langle A \rangle 
	&=& \frac{1}{\sum_{xnn'}\rho(x,n,n')} \sum_{xnn'}\rho(x,n,n')\frac{\langle x|AT^{(n)}|\phi \rangle}{\langle x|T^{(n)}|\phi \rangle},
	\label{eq:projectionMC}
\end{eqnarray}
where
\begin{eqnarray}
	\rho(x,n,n') = w_{n}w_{n'}\langle \phi |T^{(n')}|x\rangle \langle x|T^{(n)}|\phi \rangle.
\end{eqnarray}
Using the Monte-Carlo method for the summation $\sum_{xnn'}$ may reduce the computational cost.
Since $\rho(x,n,n')$ is a complex number, we need to generate $(x,n,n')$ with probability proportional to $|\rho(x,n,n')|$ and reweight it with the factor $\rho(x,n,n')/|\rho(x,n,n')|$.
That is, we calculate two quantities
\begin{eqnarray}
\alpha &=& \frac{\sum_{xnn'}\rho(x,n,n')}{\sum_{xnn'}|\rho(x,n,n')|}, \notag \\
\beta &=& \frac{1}{\sum_{xnn'}|\rho(x,n,n')|} \sum_{xnn'}\rho(x,n,n')\frac{\langle x|AT^{(n)}|\phi \rangle}{\langle x|T^{(n)}|\phi \rangle},
\end{eqnarray}
and the ratio $\beta/\alpha$ gives the value of Eq.(\ref{eq:projectionMC}).

Our improvement of the VMC method opens a way for studying systems with
various kinds of competing phases under the competition of electron correlations and spin-orbit interaction.
The Kitaev model, studied in this paper, is a good example.
The present method has a plenty of flexibilities and is straightforwardly applicable to more realistic but complicated systems including the itinerancy of electrons coexisting with electron correlations and strong spin-orbit interactions.  
Studies on the Kitaev spin liquid in more realistic cases\cite{Jackeli,Chaloupka,Kimchi,Katukuri,Rau2014,Yamaji} is
 intriguing in terms of the application of our method.
 %\cyan{[If you want to cite Kimchi, we should
%cite Katukuri {\it et al}.\cite{Katukuri}, Rau {\it et al}\cite{Rau2014}. also]}.

\section*{Acknowledgement}

The authors thank financial support by Grant-in-Aid for Scientific Research 
(No. 22340090), from MEXT, Japan.
The authors thank T. Misawa and D. Tahara for fruitful discussions.
A part of this
research was supported by the Strategic Programs for Innovative
Research (SPIRE), MEXT (grant numbers hp130007 and hp140215), and the Computational Materials Science
Initiative (CMSI), Japan.

\appendix
\section{Point Group Symmetry}
By using our VMC method, we can calculate the wave function transformed by operators of point group symmetry.
For example, in the case of hexagonal 24 sites configuration, we find that the representation with basis set defined by Eq. (\ref{xyzbasis}) reduces to $A_{1g} \oplus E_{g}$.
This is clarified by observing transformation property of the basis by $D_{3d}$ elements using Wilson loop projections and Pfaffian calculations.
Here, transformation of the function is done by using the definition of the point group elements
\begin{eqnarray}
Tc_{I\sigma}T^{-1} = a(I,\sigma,T,\uparrow)c_{T(I)\uparrow}^{\dagger} + a(I,\sigma,T,\downarrow)c_{T(I)\downarrow}^{\dagger}
\end{eqnarray}
and Eq.(\ref{eq:Fijtrans}), where $T(I)$ is the site obtained after transformation $T$ to site $I$.

\section{Exact Energy and Degeneracy at Large Sizes}
\begin{figure}[bt]
\centering
\includegraphics[width=8.5cm]{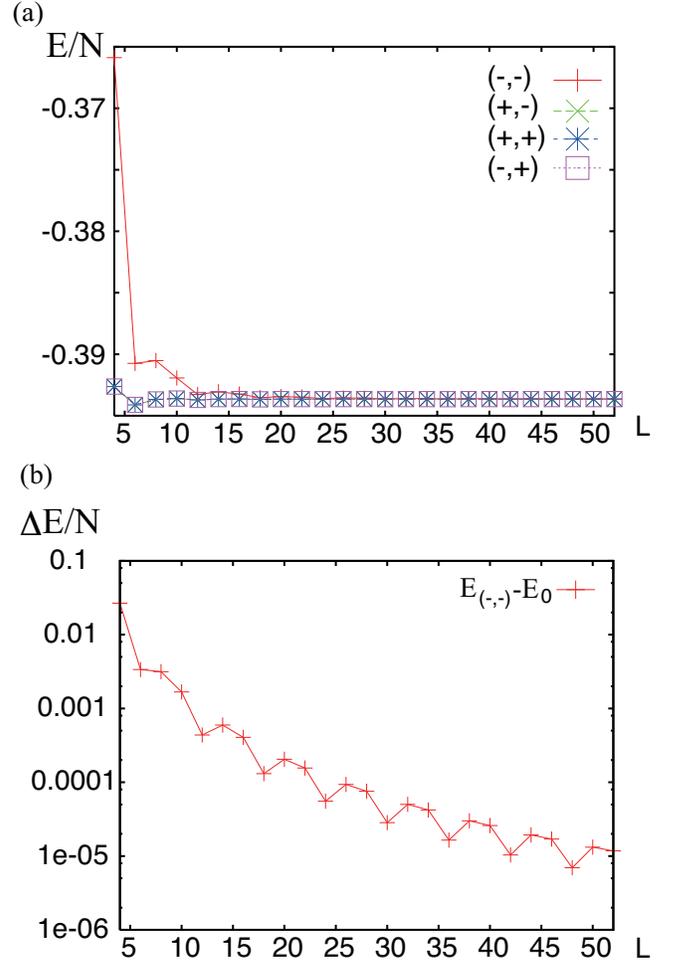}
\caption{(Color online)
(a) Exact lowest energies per site of vortex free states
for the Kitaev model with $L\times L$ periodic boundary condition.
Within the vortex free sector, the lowest energies of states distinguished
by the quantum number for the global Wilson loop operators, $W_x$ and $W_y$ are all plotted.
%The inset shows four lowest energy eigenstates in the vortex free sector.
The pairs of signs in the brackets in the legend correspond to the sign of $(W_x, W_y)$.
When $L$ is even, only $(-,-)$ state is excited and the other three states are degenerate and the ground states.
(b) Excitation energy per site of the first excited states characterized by $(-,-)$ measured from the ground state energy.
%[Should be measured from the ground state energy for each $L$ and use semi-log plot (log $y$).
%However, please keep the original figure as an inset.]
%[Add the label of the vertical axis.]
}
\label{fig:fig8}
\end{figure}
Here, we show that exact ground states of the Kitaev model have four-fold degeneracy characterized by different eigenvalues of global Wilson loop operators at large size limit.
We obtained exact solutions using the projection onto the physical solution of the Kitaev's mapping to Majorana fermions\cite{Pedrocchi}.
Figure \ref{fig:fig8} shows the lowest energy per site of vortex free states ($K_q=1$ for all hexagons) characterized by different eigenvalues of global Wilson loop operators at $L \times L$ unit cells model.
The size $L=4$ corresponds to the $4 \times 4$ unit-cell configuration.

\end{document}